# Rise and fall of defect induced ferromagnetism in SiC single crystals

Lin Li[a,b], S. Prucnal[b], S. D. Yao[a], K. Potzger[b], W. Anwand[c], A. Wagner[c], and Shengqiang Zhou[b]

[a]State Key Laboratory of Nuclear Physics and Technology, School of Physics, Peking University, Beijing 100871, China

[b]Institute of Ion Beam Physics and Materials Research, Helmholtz-Zentrum Dresden-Rossendorf, POB 51 01 19, 01314 Dresden, Germany

[c]Institute of Radiation Physics, Helmholtz-Zentrum Dresden-Rossendorf, POB 51 01 19, 01314 Dresden, Germany

*6H*-SiC (silicon carbide) single crystals containing $V_{Si}$-$V_C$ divacancies are investigated with respect to magnetic and structural properties. We found that an initial increase of structural disorder leads to pronounced ferromagnetic properties at room temperature. Further introduction of disorder lowers the saturation magnetization and is accompanied with the onset of lattice amorphization. Close to the threshold of full amorphization, also divacancy clusters are formed and the saturation magnetization nearly drops to zero.

§

Ferromagnetism in nominally diamagnetic materials such as H irradiated graphite [1], Li doped ZnO [2], defective $TiO_2$ [3], or SiC [4] recently entered the focus of research. While in the first two examples defect induced ferromagnetism is related to an extrinsic impurity, the latter two are related to intrinsic point defects like vacancies. Those can be achieved either by adjusting the growth/annealing conditions or by particle bombardment. Spin-polarized charge carriers for spin injection, magneto-optical effects, or manipulation of magnetism by external electric fields [5] are application-relevant effects expected in those materials. Note that the microscopic origin of both the magnetic moments as well as the exchange interaction is not clear in most defective ferromagnetic materials. In contrast, in Ref. [1] it was shown that the magnetic order originates solely from the carbon π-electron system. In Ref. [4], on the other hand, it was shown that in neutron bombarded *6H*-SiC single crystals, $V_{Si}$-$V_C$ divacancies form local magnetic moments arising from sp states which couple ferromagnetically due to the extended tails of the defect wave functions. While thermal neutron bombardment is a well-suited method for the creation of divacancies within the whole crystal, for applications a less expensive method to create a defined ferromagnetic film is required. Moreover, the location of ferromagnetic defects should be adjustable, as in the case of ion implantation with defined energy. In this paper we show that pronounced ferromagnetic properties of undoped *6H*-SiC single crystals can be achieved by irradiation with energetic noble-gas ions serving as chemically neutral triggers for the defect. Moreover, it is shown that the ferromagnetic order breaks down with increasing lattice disorder due to increasing ion fluence.

§

Structural defects in ion irradiated *6H*-SiC single crystals have been studied for long time. E.g. in Ref. [6] it was found that *6H*-SiC (0001) single crystals irradiated with 800 keV Ne$^+$ ions exhibit a structural transition from the crystalline to the amorphous phase. The critical number of irradiation induced displacements per atom (dpa) for complete amorphization amounts to ~0.5 dpa at 300 K as was found by Rutherford backscattering and Raman spectroscopy. This value is similar to those obtained in the same reference for irradiation with other chemically neutral ions like Ar$^+$ (1 MeV), Ar$^{2+}$ (360 keV) and Xe$^+$ (1.5 MeV) which are in the range between 0.4 and 0.5 dpa which demonstrates the high reproducibility of the structural transition. At larger fluences, the thickness of the disordered layer increases. On the other hand, ion implantation generates point-like defects in the crystalline lattice. Using positron annihilation spectroscopy it was found that the $V_{Si}$-$V_C$ divacancy, as observed in Ref. [4], is a stable defect which appears under irradiation with different elements such as Ge [7] or He [8]. Defect agglomerations along with amorphization could be observed with increasing ion fluence [7].

§

For the experiment, semi-insulating one-side polished *6H*-SiC (0001) single crystals from the KMT Corporation (Hefei, China) have been implanted with Ne$^+$ ions at an energy of 140 keV. The fluences implanted along with the induced maximum dpa calculated using SRIM-code [9] are listed in table I. The fluences have been chosen to not exceed the amorphization threshold of ~0.5 dpa specified in Ref. [6]. All samples have been implanted at room temperature in order to avoid annealing of the ferromagnetically active defects. The implantation angle was 7° with respect to the surface normal in order to reduce the channelling effect. The projected range of the ions amounts to 185 nm with a straggling of 49 nm [9].

§

Ferromagnetic hysteresis loops have been recorded at 5 K as well as 300 K using a MPMS-XL magnetometer from Quantum Design. The inset of Fig. 1a shows the magnetometry data measured at 5 K for all samples without subtraction of the diamagnetic background. The virgin material shows a marginal ferromagnetic hysteresis (Fig. 1a, inset) which has not been subtracted for the implanted samples. The magnetization M[sample] has been related to the mass of each sample. For comparison, Fig. 1a (main frame) exemplarily displays the ferromagnetic hysteresis loop for sample 1X14, i.e. the one with the largest saturation magnetization $M_S$ at both 5 K and 300 K. The diamagnetic background has been subtracted and M[layer] was now related to the mass of a thin film of 460 nm thickness which represents the total thickness of the defective layer obtained from slow positron implantation spectroscopy (SPIS) explained below. The evolution of $M_S$[layer] measured at both 5K and 300 K with increasing fluence is displayed in Fig. 1b. There are several differences as compared to the magnetic data from Ref. [4]. First of all, from the inset in Fig. 1a there is only a small increase of the slope of the linear background with increasing $M_S$ in sharp contrast to Ref. [4]. Consequently, in the ion implanted samples the majority of the defect induced magnetic moments are involved in the ferromagnetic coupling and only a small amount of them is paramagnetic. This hints towards a thin ferromagnetic defective layer defined by the implantation energy rather than a distribution of the moments over the whole sample volume. Second, the hysteresis loops measured at 5 K show larger coercive fields, e.g. due to size effects. Finally, $M_S$ drops again with increasing fluence. The origin of this behaviour is discussed along with structural investigations as follows.

§

For structural characterization, μ-Raman spectroscopy using a Nd:YAG Laser with 532 nm wavelength recorded in the scattering geometry with a liquid nitrogen cooled charge-coupled device camera has been performed. Fig. 1c exemplarily displays the Raman spectra for samples Virgin, 1X14 and 1X15, respectively. With increasing fluence, the height of the SiC related peaks is reduced, as observed in Ref. [4]. This behaviour reflects the increasing disorder of the crystalline material with increasing fluence applied. As expected, the relative intensity variation of the folded longitudinal optical (FLO) mode given by $1-I/I_0$ (Fig. 1d) nearly reaches saturation for the largest $Ne^+$ fluence implanted ($I_0$ represents the maximum intensity for the virgin sample). The saturation is associated with the threshold for complete amorphization at ~0.5 dpa as quantified in Ref. [6]. This is in contrast to Ref. [4] where no saturation was observed.

§

For the investigation of the $V_{Si}$-$V_C$ divacancies, monoenergetic SPIS has been performed recording the S-parameter versus the incident positron energy. In brief, the momentum of the electron–positron pair prior to annihilation causes a Doppler broadening of the 511 keV annihilation line and can be characterized by the line-shape parameter S. The value of S is defined by the ratio of counts in the central region of the annihilation gamma peak to the total number of counts in the peak. The energy regions are selected to give S=0.5 for the virgin bulk material and then fixed for the irradiated samples. The S-parameter characterizes the positron annihilation with low momentum electrons while it increases with increasing size of the particular open volume defects (e.g. defect clustering). The variation of the incident positron energy enables the measurement of a depth dependence of S. For a more general discussion of SPIS we refer to Refs. [10-11]. In Ref. [7] it was shown that the $V_{Si}$-$V_C$ divacancy in *6H*-SiC can be identified by a value of the S-parameter of 0.5375. In Ref. [12], a relationship between the number of clustered divacancies and the S-parameter was established. Fig. 1e displays the dependence of the S-parameter associated with the $V_{Si}$-$V_C$ divacancy on the implantation energy of the positrons and their mean penetration depth. It is evident that the defective region containing divacancies is a thin layer in the near surface region of the *6H*-SiC single crystal. An estimation of the thickness d of the damaged layer containing $V_{Si}$-$V_C$ divacancies from the SPIS data resulted in d=460±25 nm. As depicted in Fig. 1e, the S-parameter drastically increases between a fluence of 0.24 and 0.48 dpa reaching a maximum value of 0.55 (=1.1 $S_{bulk}$, , where $S_{bulk}$ represents the defect-free material) for sample 1X15 at a mean positron penetration depth of 120 nm. For comparison, Fig. 1f displays the evolution of the S-parameter with respect to $S_{bulk}$ at a penetration depth of 120 nm as well as the associated number of agglomerated divacancies obtained from Ref. [12] along with the dpa. Surprisingly, the S-parameters of samples 1X14 and 5X14 nearly overlap despite their pronounced difference of $M_S$. Thus, the drop of $M_S$ is not primarily connected to an increase of the S-parameter, i.e. not to the change of the size of the largest open volumen defects. While for nearly all of the samples a single divacancy was obtained, for sample 1X15, 5 agglomerated $V_{Si}$-$V_C$ divacancies are the major defects. Those are, according to Raman spectroscopy (Fig. 1c-d), embedded in an amorphous matrix. Note that the actual maximum concentration of divacancies given by half of the dpa is much reduced due to self annealing effects. Within the sensitivity of the experiment, a rough estimation of the divacancy concentration using SPIS revealed a reduction by a factor of at most 120. Consequently, the mean divacancy distances ranging from approximately 4.7 nm for sample 5X13 to 2.2 nm for sample 5X14 are in reach of the value of 1.23 nm specified in Ref. [4].

§

Comparing the above mentioned results, the initial increase of the saturation magnetization with increasing fluence (Fig. 1b) can be related to an increase of the density of $V_{Si}$-$V_C$ divacancies and thus a higher density of the magnetic moments. This observation is consistent with Ref. [4], where an increase of $M_S$ with increasing neutron fluence was observed. The decrease of $M_S$ at a fluence of 0.24 dpa (sample 5X14) with respect to 0.048 dpa (sample 1X14) may have different sources. The discussion of a possible change of the magnetic coupling from ferromagnetic to antiferromagnetic [4] is beyond the scope of the paper. Instead, we would like to highlight structural changes occurring with increasing ion fluence possibly affecting the magnetic properties, i.e. the onset of amorphization. Already a weak degeneration of the crystalline structure might lead to the suppression of ferromagnetic exchange as in the case of Mn-doped GaAs [13]. In Ref. [6] it was found that at 300 K irradiation temperature, detectable Si disorder occurs at around 20% of the fluence necessary for total amorphization. In our case, this value corresponds to twice the fluence applied to sample 1X14 explaining the decrease of $M_S$ for samples 5X14 and 1X15. It was also shown in Ref. [6] that the disorder detected by Raman spectroscopy as in our case is connected directly to the onset of stable amorphous areas in the material. Thus, amorphization introduced by the implantation reduces the thickness of the ferromagnetic layer. At a fluence of $1\times10^{15}$ cm$^{-2}$ corresponding to 0.48 dpa the material is nearly fully amorphized and the divacancies are forming agglomerates.

§

Concerning electronic transport properties and comparing samples virgin and 1X14, we did not observe an insulator-to metal transition as expected for common diluted magnetic semiconductors. Both samples are semi-insulating with a sheet resistance above 20 M$\Omega$. An ion irradiation induced conductivity increase can also not be expected since it can even lead to resistivity increase. In Ref. 14, it was found that for 70 keV oxygen implanted n-doped *6H*-SiC, a highly resistive layer is formed at fluences of $1\times10^{14}$ cm$^{-2}$ (0.072 dpa) to $5\times10^{14}$ cm$^{-2}$ (0.36 dpa). The dpa values are indeed comparable to those for samples 1X14, 5X14 and 1X15 (Table I). The mechanism for magnetic ordering in defective SiC thus should rather be compared to other insulating diluted magnetic compounds such as Cr doped $TiO_2$ [15].

§

In conclusion we found that Ne$^+$ ion irradiation of *6H*-SiC single crystals leads to the formation of defect induced room-temperature ferromagnetism which is considerably larger as for the case of neutron irradiated material. The saturation magnetization increases initially with increasing fluence but drops with increasing disorder. Finally we would like to remind that SiC is a common substrate for semiconductor thin films. Thus, unwanted defect induced ferromagnetism in SiC could be wrongly associated to the deposited film which might lead to data misinterpretation.

§

L.L. thanks the financial support by China Scholarship Council (File No. 2009601260) for her stay at HZDR and by the National Basic Research Program of China (Grant No. 2010CB832904). S.Z. acknowledges funding by the Helmholtz-Gemeinschaft Deutscher Forschungszentren (HGF-VH-NG-713).

Figure captions

Fig. 1(a). Ferromagnetic hysteresis loops recorded at 5 K or 300 K for sample 1X14. The magnetization M was related to a thin layer of 460 nm thickness. The 5 K as-measured loops without subtracting the diamagnetic background are shown in the inset. The magnetization M was related to the whole sample weight. (b) Evolution of the saturation magnetization with the displacements per atom (dpa). (c) Raman spectra for virgin and $Ne^+$ implanted *6H*-SiC single crystals. (d) shows the relative intensity variation of the folded longitudinal optical (FLO) mode marked in (c) along with the displacements per atom (dpa). (e) S-parameter depending on positron implantation energy and $Ne^+$ fluence implanted. Since the positron implantation profile corresponds to a broad Makhov-type distribution [12], it probes defects also with the distribution tails, i.e. the mean positron implantation depth does not correspond directly to the geometric depth of the sample. (f) Dependence of the S-parameter (with respect to the bulk value $S_{bulk}$) as well as the estimated number N(agglomerated) of on the $V_{Si}$-$V_C$ divacancies in the cluster on the displacements per atom (dpa).

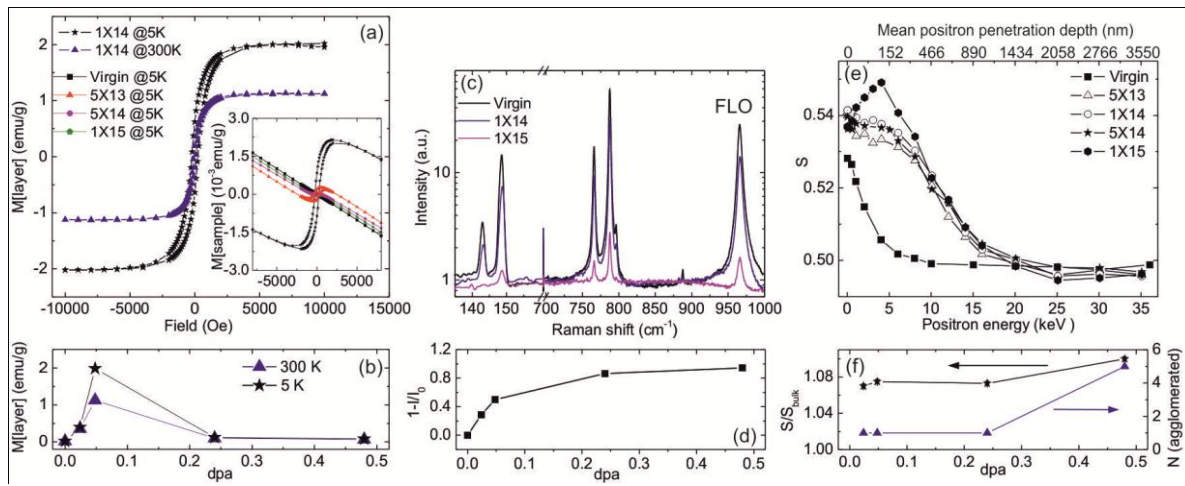

Table I. Sample identifiers, ion energy, ion fluences and maximum dpa of *6H*-SiC single crystals implanted with Ne$^+$ ions.

| Sample identifier | Energy (keV) | Fluence (cm$^{-2}$) | Maximum displacements per atom (dpa) |
|---|---|---|---|
| Virgin | - | - | - |
| 5X13 | 140 | 5x10$^{13}$ | 0.024 |
| 1X14 | 140 | 1x10$^{14}$ | 0.048 |
| 5X14 | 140 | 5x10$^{14}$ | 0.24 |
| 1X15 | 140 | 1x10$^{15}$ | 0.48 |